\documentclass[twocolumn,showpacs,preprintnumbers,amssymb]{revtex4}

\usepackage{graphics}
\usepackage{epsfig} 
\usepackage{hyperref}
\usepackage{graphics}
\usepackage{color}      % use if color is used in text
\usepackage{graphicx}
\usepackage{graphicx}% Include figure files
\usepackage{dcolumn}% Align table columns on decimal point

\usepackage{epstopdf}
\usepackage{color}
\usepackage{dcolumn}
\usepackage{amsmath,amssymb}

\usepackage{amsmath}

\begin{document}

 % 1 rescaled parameters for the figure
% 2.put more reference and correct  references
% 3. go through out the paper and edit.
%4. print out and edit (3 times)
%5. with god wishs submit to pre

\title { Thermodynamic Irreversibility in Underdamped Brownian Motion with Spatial Temperature Gradients}
\author{Mesfin Asfaw  Taye}
\affiliation {West Los Angles College, Science Division \\9000  Overland Ave, Culver City, CA 90230, USA}%Lines break automatically or can be forced with \\

\email{tayem@wlac.edu}

\begin{abstract}

In this work,    we extensively    explore  the thermodynamic properties of Brownian particles moving in an underdamped medium. Extending  our   previous analysis \cite{muuu177}, we examine systems subjected to different thermal arrangements,  such as    quadratically and linearly decreasing temperature profiles, as well as piecewise constant temperature distributions. By conducting rigorous  derivations, we derived     several  thermodynamic relations  that  provide  insights into non-equilibrium thermodynamics and the underlying mechanisms governing entropy production and extraction. Furthermore, we address a fundamental question that includes whether the vanishing entropy production or extraction rate necessarily implies a thermodynamic equilibrium. Our analytical findings reveal that for a Brownian particle in an underdamped medium (free from external forces or ratchet potentials), both the entropy production and extraction rates decrease to zero, even in the presence of a spatially varying temperature gradient. However, the total entropy production ($E_P > 0$) and total entropy extraction ($H_d > 0$) remain finite, indicating that the system retains intrinsic irreversibility driven by heat exchange via kinetic energy transfer. This also  implies that 
 the zero-entropy   production rate  does not signify  an equilibrium.    Our results further indicate that in the absence of external loads and potentials, most thermodynamic rates asymptotically decay to zero.  However, the system’s irreversibility persists because of the continuous heat flow from regions of higher to lower temperatures.  We also consider the motion of  a  Brownian particle that moves  in a ratchet potential coupled with a spatially varying temperature. Such particles move unidirectionally,  even in the absence of force.  By imposing asymmetry in the thermal arrangements,  one can manipulate the direction of the particles,  indicating that  this  finding has significant applications in microscale and nanoscale transport systems.  We show that, because the velocity of a particle depends on its mass,    barrier height, load, and  noise intensity, the particles can be sorted along the reaction coordinate  depending on their physical properties.
\end{abstract}
\pacs{Valid PACS appear here}% PACS, the Physics and Astronomy
\maketitle

%\section{Introduction}

%%%

 \section{Introduction}

 The thermodynamics of systems far from equilibrium are difficult to analyze because their macroscopic quantities often depend on microscopic transition rates in complex ways. This contrasts with equilibrium systems, where thermodynamic relationships are well understood and clearly defined. Recent advances based on Boltzmann-Gibbs entropy and the entropy balance equation \cite{mu1,mu2,mu3,mu4,mu5,mg11,mu6,mu7,mu8,mu9,mu10,mu11,mu12,ta1,mu13,mu14,mu15,mu16} have provided strong theoretical tools to study such systems. Solvable models \cite{mu17,muu17} are able to  explain how entropy production and extraction  rates change over time  as well as  how they are influenced by system parameters that govern dissipation and irreversibility.  In quantum and biological systems, parameter-dependent thermodynamic behavior was studied in \cite{mu25,mu26,mu27} along  with empirical support from intracellular transport studies \cite{mu28,mu29,mu30}. More generally, the dependence  of  entropy production, extraction, free energy and other thermodynamic relations   has been explored   for Brownian particles in  overdamped \cite{muuu17} and underdamped \cite{muuu177,c1,c2,c3,c4,c5,c6,c7} regimes.   Earlier  Seifert \textit{et al.} \cite{mu6}  introduced a
 methodology for calculating the entropy production and
 extraction rates at the ensemble level by first analyzing the thermodynamic relations at the trajectory level.  Alternatively,  the non-equlibirum of the system can be studied  using  time-reversal operations  \cite{mar2,mar1}.  Recent developments in nonequilibrium thermodynamics have also explored entropy production, dissipation, and transport processes in both classical and quantum systems \cite{mg1,mg2,mg3}. These advances include theoretical frameworks, simulation methods, and experimental insights into active matter, quantum engines, and phase transitions \cite{mg4,mg5,mg6}.  In our recent study \cite{mg11}, we explored the thermodynamic behavior of a Brownian particle in an underdamped medium. Here, we extend that work by examining how different temperature profiles, such as quadratically and linearly decreasing and piecewise constant distributions, affect nonequilibrium thermodynamics along the reaction coordinate.

Does zero entropy production or extraction rate necessarily mean that a system is at equilibrium? Are there better indicators that more accurately reflect non-equilibrium conditions? In this study, we   address  whether  a vanishing entropy production or extraction rate necessarily implies that the system has attained equilibrium. By analyzing the  influence of various thermal configurations, we further studied the model system.  We show that  even in the presence of  a spatially varying temperature gradient along the reaction coordinate, the entropy production rate and entropy extraction rate decrease to zero in the absence of a force  for a Brownian particle that  moves  in an underdamped medium.   However, our  study indicates that  the total entropy production ($E_P > 0$) and total entropy extraction ($H_d > 0$) remain finite. This indicates that   the system retains its inherent irreversibility, which is driven by heat exchange through kinetic energy transfer. These findings show  that a zero-entropy production rate does not necessarily signify equilibrium, but  it highlights that the system may be  driven out of equilibrium under nonequilibrium steady-state conditions. Moreover, in  this  theoretical work, we  investigated  the dependence of heat exchange via kinetic energy  on model parameters, such as temperature gradients and viscous friction.    Importantly, our analysis reveals that a vanishing entropy production or extraction rate does not necessarily indicate thermodynamic equilibrium. Even in the absence of external forces, systems with spatial temperature gradients continue to exhibit irreversible behavior due to kinetic energy-driven heat flow. This distinction between rate-based and integrated measures of irreversibility plays a central role in our results. In what follows, we will show analytically that although the entropy production and extraction rates vanish in the zero-force limit, the total entropy production and heat dissipation remain finite—demonstrating that the system resides in a nonequilibrium steady state.

 The results  obtained in this study also indicate  that  a  Brownian particle moving in a ratchet potential coupled with a spatially varying temperature exhibits  directed motion even in the absence of an external load. The asymmetry in  the thermal arrangement, as well as the presence of load,  leads to a net particle current,  indicating that  this  finding has significant applications in microscale and nanoscale transport systems, where controlled particle movement is essential.    Moreover, the dependence of the velocity on the particle mass suggests potential applications in particle separation  based on physical properties.   The ability to control a particle's velocity  by varying  the system parameters indicates that  one can enhance  the  efficiency  of microfluidic devices and nanoscale sensors.   Overall, our study  shows that the desired particle velocity  can be obtained  by varying the system parameters, such as  the barrier height, load, and  noise intensity across the reaction coordinate. We believe that these results contribute to a deeper understanding of thermally driven ratchets and highlight the role of temperature asymmetry in Brownian motors.

The remainder of this paper is organized as follows.  In Section II,  we introduce the model system and provide a rigorous derivation of entropy production and free energy expressions. In Section III, we investigate the dependence of entropy production, entropy extraction, and free energy rates on the model parameters  for a Brownian particle freely diffusing in an isothermal underdamped medium. In  Section IV, a more  rigorous  approach is presented for  linearly decreasing thermal arrangements. We then study   the thermodynamic features  for a Brownian particle moving in ratchet potential in Section V.  The role of  multiplicative noise  and viscous friction  on the entropy production and extraction rate is  studied in Section VI.  Section VII  presents the summary and conclusions.

\section{Derivation of free energy for underdamped case  }  

To account for possible external driving, we include a linear potential term \(fx\) in the total potential \(U(x) = U_s(x) + fx\), where \(f\) represents a constant external force. This formulation allows us to explore the system’s thermodynamic response across both driven and purely thermally induced regimes.

For the Brownian particle that  walks in  underdamped and isothermal medium,  the expressions for  the  entropy production and entropy extraction rates  have been derived  by 
Ta et al.  \cite{ta1} in terms of particle velocity and probability distribution. To account for possible external driving, we include a linear potential term \(fx\) in the total potential \(U(x) = U_s(x) + fx\), where \(f\) represents a constant external force and $ U_s(x)$ represents the periodic potential. This formulation allows us to explore the system’s thermodynamic response across both driven and purely thermally induced regimes.
  The dynamics of the Brownian particle can be studied via  the Langevin equation
\begin{equation}
m{dv\over dt} = -\gamma{dx\over dt}- {d U(x) \over dx}  + \sqrt{2k_{B}\gamma T(x)}\xi(t),
\end{equation}
$k_{B}$  denotes the Boltzmann constant, which is  considered to have a value of unity in this scenario. The random noise term $\xi(t)$ is assumed to follow a Gaussian white noise distribution and satisfies the conditions $\left\langle \xi(t) \right\rangle = 0$ and $\left\langle \xi(t) \xi(t') \right\rangle = \delta(t-t')$.  $\gamma$ denotes the viscous friction. The spatially varying temperature profiles $T(x)$ considered in this study are motivated by their role as canonical nonequilibrium driving mechanisms in stochastic thermodynamics. Such gradients naturally arise in micro- and nanoscale systems—such as thermophoretic transport, optical heating of colloids, or intracellular environments—where thermal inhomogeneity induces irreversibility even in the absence of external forces.
 For  the underdamped Langevin case, as discussed by  Sancho et al. \cite{am3} and Jayannavar et al. \cite{am33}, neither the Ito interpretation nor the Stratonovich interpretation is required.

Using  the Fokker-Planck equation  \begin{eqnarray} 
 {\partial P\over \partial t}&=&-{\partial (vP) \over \partial x}-{1 \over m}{\partial(U'(x)P) \over \partial v}+ \nonumber \\  &&{\gamma(x) \over m}{\partial (vP) \over \partial v}+{\gamma(x) T(x) \over m^2}{\partial^2 P \over \partial v^2},  
\end{eqnarray}  
the model dynamics can then be studied.  
Here,   $P(x,v,t)$ denotes the probability of locating a particle at a specific position $x$, velocity  $v$ and  time $t$.  One can  write Gibbs entropy as  
\begin{eqnarray}  S(t)= -\int P(x,v,t)\ln P(x,v,t) dxdv. 
 \end{eqnarray} 
 Via the method  presented in \cite{mu7}, the entropy production and dissipation rates can be derived from  the time derivative of the Gibbs entropy  as 
 \begin{eqnarray}  {d S(t)\over dt}&=& -k_{B}\int {\partial P(x,v, t)\over \partial t} \ln[P(x,v,t)]dxdv.  \end{eqnarray} 
 Eq. (4) can be rewritten as    
\begin{eqnarray}  {d S(t)\over dt}&=&{\dot e}_{p}-{\dot h}_{d} 
  \end{eqnarray} 
 where   the entropy production rate  ($\dot{e}_p$)  represents the rate of irreversible entropy production  within the system. On the other hand, the entropy extraction rate $\dot{h}_d$ denotes  the rate at which entropy is transferred to the environment via dissipative processes such as heat flow. These two rates are connected via the entropy balance equation, $dS/dt = \dot{e}_p - \dot{h}_d$. At thermodynamic equilibrium, both rates vanish: $\dot{e}_p = \dot{h}_d = 0$. In contrast, in a nonequilibrium steady state, the system entropy $S(t)$ remains constant ($dS/dt = 0$), but $\dot{e}_p = \dot{h}_d > 0$, indicating continual microscopic irreversibility. This framework is consistently used  throughout our analysis to interpret energy dissipation and thermodynamic irreversibility under various thermal gradients.

In order to calculate ${\dot h}_{d}$, let us first  find the heat dissipation rate ${\dot H}_{d}$   via stochastic energetics that discussed in the works \cite{am4,am5}. 
To derive the expression for the heat dissipation rate \( \dot{H}_d \), we begin with the underdamped Langevin equation:
$
m \frac{dv}{dt} = -\gamma(x) v - \frac{dU(x)}{dx} + \sqrt{2k_B \gamma(x) T(x)}\,\xi(t),
$
where \( \xi(t) \) is a Gaussian white noise with zero mean. Multiplying both sides by the velocity \( v = \dot{x} \), we obtain the instantaneous power balance:
$
m v \frac{dv}{dt} = -\gamma(x) v^2 - v \frac{dU(x)}{dx} + \sqrt{2k_B \gamma(x) T(x)}\,\xi(t) v.
$
Taking the ensemble average and noting that the noise term averages to zero due to \( \langle \xi(t) \rangle = 0 \), we have:
$
\left\langle m v \frac{dv}{dt} \right\rangle = - \left\langle \gamma(x) v^2 \right\rangle - \left\langle v \frac{dU(x)}{dx} \right\rangle.
$
On the other hand, the rate of heat dissipation from the particle to the environment can be defined, following the stochastic energetics framework developed by Sekimoto~\cite{am4,am5}, as:
$
\dot{H}_d = \left\langle \left( -\gamma(x) v + \sqrt{2k_B \gamma(x) T(x)}\,\xi(t) \right) \cdot v \right\rangle.
$
Again, the stochastic term averages to zero, leaving:
$
\dot{H}_d = - \left\langle \gamma(x) v^2 \right\rangle.
$
Substituting from the earlier identity, we then obtain:
\begin{eqnarray}
{\dot H}_{d}  
&=&-\left\langle \left(-\gamma(x){\dot x}+ \sqrt{2k_{B}\gamma(x) T(x)}\right).{\dot x}\right\rangle  \nonumber \\
&=&-\left\langle m{vdv\over dt}  +v U'(x)   \right\rangle,
\end{eqnarray}
Once the energy dissipation rate is obtained,  
based on our previous works \cite{mu17,muu17,muuu17},    the entropy extraction rate ${\dot h}_{d}$ then can be found as 
\begin{eqnarray}
{\dot h}_{d}  
&=&-\int \left({m{vdv\over dt}  +v U'(x) \over T(x)} \right)P dxdv.
\end{eqnarray}
At this point,  we want to stress that Eq. (6) is exact and does not depend on any boundary condition  (as can be seen in  the next sections).
Since ${d S(t)\over dt}$ and ${\dot h}_{d}  $ are computable,  the entropy production rate  can be readily  obtained as 
\begin{eqnarray}
{\dot e}_{p}&=&{d S(t)\over dt}+{\dot h}_{d}. 
\end{eqnarray}
At steady state, ${d S(t)\over dt}=0$ which implies that  ${\dot e}_{p}={\dot h}_{d}>0$. For the isothermal case, in the stationary state (approaching equilibrium), ${\dot e}_{p}={\dot h}_{d}=0$. 

Note that in our analysis, the underdamped Langevin equation is interpreted in the \textit{Itō} sense, as is standard in stochastic energetics for systems with multiplicative noise where the noise amplitude depends on position but not on velocity. This choice is crucial in evaluating ensemble averages involving stochastic terms. In particular, we use the identity $\langle \xi(t) v(t) \rangle = 0$, which holds under the Itō convention. While the Fokker--Planck equation (Eq.~(2)) is invariant under different stochastic interpretations, thermodynamic quantities such as the heat dissipation rate are interpretation-sensitive. If one instead adopts the Stratonovich framework, additional drift corrections (e.g., via the Furutsu--Novikov formula) must be included. However, our use of the Itō prescription ensures mathematical consistency with the expressions derived in Eqs.~(6) and (7), and the conclusions of our work remain robust under this choice.

In addition, in situations where the probability distribution exhibits periodic behavior or diminishes at the boundary, Tome $et$. $at.$ \cite{ta1} established equations for the rates of entropy production and entropy extraction under isothermal conditions. Building upon their methodology, we can rephrase Equation (2) accordingly.
\begin{eqnarray}
{\partial P\over \partial t}&=& k+{\partial J' \over \partial v} 
\end{eqnarray}
where 
\begin{eqnarray}
k=v{\partial P \over \partial x} +{1\over m}(U'){\partial P \over \partial v}
\end{eqnarray}
and 
\begin{eqnarray}
J'= -{\gamma(x)\over m}vP-{ T(x)\over m^2}{\partial P\over \partial v}.
\end{eqnarray}
After applying a boundary condition, the variable $k$ becomes zero.
After some algebra one gets 
\begin{eqnarray}
{\dot e}_{p}&=& -\int {m^{2}J'^{2} \over P T(x) \gamma(x)} dx dv
\end{eqnarray}
and 
\begin{eqnarray}
{\dot h}_{d} &=& -\int  {mvJ' \over T(x) } dx dv
\end{eqnarray} respectively.

Because the expressions for ${\dot S}(t)$, ${\dot e}_{p}(t)$  and ${\dot h}_{d}(t)$  can be determined at any time $t$ the analytical expressions for the changes in entropy production, heat dissipation, and total entropy can be found.  
$
\Delta h_d(t)= \int_{0}^{t}{\dot h}_{d}(t)dt$, 
$\Delta e_{p}(t)= \int_{0}^{t}  {\dot e}_{p}(t)  dt$ and 
$\Delta S(t) =\int_{0}^{t} {\dot S}(t)dt$
where $\Delta S(t)=\Delta e_p(t)-\Delta h_d(t)$,

Next, we  derive the expression for the free energy dissipation rate ${\dot F}(t)$ in terms of  ${\dot E}_{p}(t)$   and  ${\dot H}_{d}(t)$. ${\dot E}_{p}(t)$   and  ${\dot H}_{d}(t)$ are the terms associated with ${\dot e}_{p}(t)$   and  ${\dot h}_{d}(t)$.   Let us 
 now introduce  ${\dot H}_{d}(t)$ for the considered model system.   The heat dissipation rate   is either  given 
\begin{eqnarray}
{\dot H}_{d} &=& -\int  {mvJ' dx dv}.
\end{eqnarray}
Equation (14) is notably different from Eqs. (7) and (13), due to the  term $T(x)$. 
On the other hand, the term  ${\dot E}_{p}$ is  related to  ${\dot e}_{p}$ and it is given by 
\begin{eqnarray}
{\dot E}_{p}&=& -\int {m^{2} J'^{2} \over P \gamma(x)} dx dv.
\end{eqnarray}
The new entropy balance equation  
\begin{eqnarray}
{d S^T(t)\over dt}&=&{\dot E}_{p}-{\dot H}_{d},
\end{eqnarray}

Since the expressions for ${\dot S}^T(t)$, ${\dot E}_{p}(t)$  and ${\dot H}_{d}(t)$
 can be obtained as functions of time $t$, the analytical expressions for the changes related to the rates of entropy production, heat dissipation, and total entropy can be found analytically via 
$
\Delta H_d(t)= \int_{0}^{t}{\dot H}_{d}(t)dt $
$\Delta E_{p}(t)= \int_{0}^{t} {\dot E}_{p}(t) dt $ and 
$\Delta S(t)^T =\int_{0}^{t} {\dot S}(t)^{T}dt$
where $\Delta S(t)^T=\Delta E_p(t)-\Delta H_d(t)$,

On the other hand, the internal energy is given by 
\begin{eqnarray}
{\dot E}_{in} = \int ({\dot K}+ v U'_{s}(x))P(x,v,t)dvdx,
\end{eqnarray}
where   ${\dot K}=m{vdv\over dt}$  and $U'_{s}$ denote the kinetic and potential energy rates, respectively.
 For a Brownian particle operating in a spatially varying temperature field, the total work done is given by 
\begin{eqnarray}
{\dot W}&=& \int v f P(x,v,t)dvdx,
\end{eqnarray}
The first law of thermodynamics  can be written as  
\begin{eqnarray}
{\dot E}_{in} = -{\dot H}_{d}(t)-{\dot W},
\end{eqnarray}
 The change in the internal energy   reduces to  
$
\Delta E_{in}= -\int_{0}^{t}( {\dot H}_{d}(t)+{\dot  W}) dt 
$ 
we write the  free energy dissipation rate as
\begin{eqnarray}
{\dot F}&=&{\dot E}_{in}- {\dot S}^T \nonumber \\
&=&{\dot E}_{in}-{\dot E}_{p}+{\dot H}_{d},
\end{eqnarray}
The change in the free energy is given by  
\begin{eqnarray}
\Delta F(t)&=&-\int_{0}^{t} \left(  {\dot W}+ {\dot E}_{p}(t)   \right)dt.
\end{eqnarray} 
For isothermal case, at  quasistatic limit where the velocity  approaches  zero  $ v=0$, ${\dot E}_{p}(t) =0$ and ${\dot H}_{d}(t) =0$ and far from quasistatic limit 
$E_{p}={\dot H}_{d}>0$  which is  expected as   the particle  operates irreversibly.

At this point, we emphasize that the term \( S^T \) in Eqs.~(16)  and (20) denote the total entropy-like quantity expressed in energetic units.  Although it is closely related to the system entropy \( S \) introduced in Eq.~(5), they are not identical in their form or interpretation. Specifically, Eq.~(5) represents the conventional entropy balance,  and in this case,  the time derivative of the system entropy equals the difference between the entropy production rate and the entropy extraction rate. These terms are all expressed in units of entropy and characterize the microscopic irreversibility of the dynamics. In contrast, Eqs.~(16)  and (20) reformulate the entropy balance within the framework of energy flows following the principles of stochastic energetics developed by Sekimoto~\cite{am4,am5}.  In this representation, entropy-related quantities are multiplied by local temperature fields to yield energy rates, and the resulting expression connects the free energy dissipation, internal energy changes, and an energy-based entropy-like term \( S^T \).  The definition \( \dot{S}^T \equiv \dot{E}_p - \dot{H}_d \) reflects the net energetic balance between entropy production and dissipation, where \( \dot{E}_p \) and \( \dot{H}_d \) are the energetic counterparts of entropy production and heat extraction rates, respectively, as introduced in Eqs.(14) and (15), respectively.Thus, while \( S \) and \( S^T \) are conceptually linked through their roles in nonequilibrium thermodynamics, the latter is a temperature-weighted analog that facilitates energy-based thermodynamic relations  in underdamped stochastic systems.

\section{ Entropy production and extraction rate  for  different  thermal arrangements and the heat exchange via kinetic energy}

In this section, we  show that   for the underdamped   case,  as long as a non-uniform  thermal arrangement is imposed  along the reaction coordinate,   the system  is driven out  of equilibrium  even in the absence of symmetry-breaking fields such as  external force.  In order to get  a deeper understanding,   we solve Eq. (2)  at steady state and after some algebra one gets 
\begin{equation}
P(v, x) = \sqrt{\frac{m}{2\pi T(x)}} \exp\left[ -\frac{m \gamma(x)}{2 T(x)} \left( v - \frac{f}{\gamma(x)} \right)^2 \right].
\end{equation}
While solving the above  equation,    we assume that the system is in a steady state, that the particle mass \( m \) is small, and that the boundary conditions are periodic. Under these conditions, the first term in the Fokker–Planck equation (Eq.~(2)), involving the spatial derivative \( \partial_x (vP) \), becomes negligible—either due to averaging over a full spatial period or because its contribution is small compared to velocity-space terms. Additionally, we impose that the velocity distribution is normalized to unity at each position \( x \). These assumptions enable us to obtain a tractable and physically meaningful approximate solution for the steady-state distribution.

{\it Entropy production and extraction rate  for  different  thermal arrangements.\textemdash}   We now proceed to calculate the entropy production and extraction rates for a quadratic thermal arrangement: 
\begin{equation} 
  T(x)=T_{h}+{(T_{c}-T_{h})x^2\over L_o^2}.
   \end{equation}
	Here the temperature  decreases  quadratically  from the hot bath $T_h$ to cold bath $T_c$ along the reaction  coordinate  $L_0$.  Let us assume that the viscous friction to be independent of $x$. 
  Via Eqs. (23), (12) and (13), and performing some algebraic manipulations, we derive that the entropy production and extraction rates are given by
 \begin{eqnarray} 
{\dot h}_{d}(x)&=&{\dot e}_{p}(x,t) \nonumber \\
&=& {f^{2}\over \gamma (T_{h}+{(T_{c}-T_{h})x^2\over L_o^2})}
   \end{eqnarray}
	{\it  Here  in order to investigate how entropy production and extraction behave as a function of position, we perform integration solely with respect to velocity}.
	In the limit $f \to 0$,  ${\dot h}_{d}(x)={\dot e}_{p}(x) \to 0$ as anticipated.
 The average velocity is found to be  
\begin{equation} 
%v={2fL_{0}+(T_c-T_h)\over 2\gamma L_0}.
v={f\over \gamma }.
\end{equation}
 In the absence of force, the velocity approach zero. However, this does not indicate that the system reaches an equilibrium point.  While $v=0$, the particle speed remains non-zero, and as the particle transitions from the hotter bath to the colder bath, it continuously dissipates heat via kinetic energy. 

% ---- Start of Section ----

For linearly decreasing temperature case
\begin{equation} 
  T(x)=T_{h}+{x(T_{c}-T_{h})\over L_o},
\end{equation}
employing Eqs. (26), (12) and (13), the entropy production and extraction rates are computed as follows:
\begin{eqnarray} 
\dot{h}_{d}(x)&=&\dot{e}_{p}(x) \nonumber \\
&=& {f^{2}\over \gamma \left(T_{h}+{x(T_{c}-T_{h})\over L_o} \right)}.
\end{eqnarray}
From Eqs. (24) and (27) it is evident that in the limit where the load approaches zero $f \to 0$, $\dot{h}_{d} = \dot{e}_{p} = 0$. This, however, does not imply that the system reaches equilibrium. In nonhomogeneous thermal arrangements, there is an irreversible heat flow from the hotter to the colder bath via kinetic energy. From Eqs. (24) and (27), it is apparent that in the isothermal limit where $T_h \to T_c$, the entropy dissipation rate $\dot{h}_{d}(x)$ and the entropy production rate $\dot{e}_{p}(x)$ both simplify to 
${f^{2}/(\gamma T_c)}$ as expected, since as long as the load is non-zero, the system is driven out of equilibrium. As the temperature difference between $T_h$ and $T_c$ increases, both  $\dot{h}_{d}(x)$ and $\dot{e}_{p}(x)$ increase correspondingly. 
	
The rate of heat dissipation is calculated via Eqs. (6) and (14), and it converges to
\begin{eqnarray} 
\dot{H}_{d}(x) = \frac{f^2}{\gamma}.
\end{eqnarray}
Conversely, the rate of work done is expressed as:
\begin{eqnarray} 
\dot{W}(x) = \dot{E}_{p}(x) \nonumber \\ 
= \frac{f^2}{\gamma}.
\end{eqnarray}
For the isothermal case $T_{h} = T_{c}$, one obtains $v = f/\gamma$, $\dot{h}_{d} = \dot{e}_{p} = f^2 L_{0}/(\gamma T_{c})$, and $\dot{H}_{d} = \dot{E}_{p} = f^2 L_{0}/\gamma$. Note that we integrate out $x$.

{\it The role of the heat exchange via kinetic energy.\textemdash}
In the limit where the load approaches zero, the rates of entropy production and extraction become zero. However, this does not imply that the system reaches equilibrium. As long as a non-uniform thermal arrangement exists across the reaction coordinate, $\dot{H}_{d} = \dot{E}_{p} \neq 0$. To delve deeper, we rewrite the extraction rate as:
\begin{eqnarray}
\dot{H}_{d}(x) 
&=& -\left\langle \left( -\gamma(x)\dot{x} + \sqrt{2k_{B} \gamma(x) T(x)} \right) \cdot \dot{x} \right\rangle  \nonumber \\
&=& -\left\langle m v \frac{dv}{dt} + v U'(x) \right\rangle.
\end{eqnarray}
From the above equation, it follows that $H_{d} = -\left\langle \frac{mv^2}{2} + U(x) \right\rangle$. In the absence of a ratchet potential, this expression can be further manipulated as:
\begin{eqnarray}
H_{d}(x) &=& -\int_{-\infty}^{\infty}\left( m v dv + U'(x) dx \right) P(x,v) \nonumber \\
&=& -\frac{1}{2}\left(2f x + \frac{f^2 m}{\gamma^2} + T(x) \right).
\end{eqnarray}
In the absence of load, 
\begin{eqnarray}
H_{d}(x) &=& -\int_{-\infty}^{\infty} \left( m v dv + U'(x) dx \right) P(x,v) \nonumber \\
&=& -\frac{T(x)}{2}.
\end{eqnarray}
This implies that the change in the internal energy reduces to  
\begin{eqnarray}
\Delta E_{\text{in}}(x) &=& -\int_{0}^{t} \dot{H}_{d} \, dt \nonumber \\
&=& \frac{T(x)}{2}.
\end{eqnarray}

The change in the free energy (in the absence of load) is given by  
\begin{eqnarray}
F(x) &=& -\int_{0}^{t} \dot{E}_{p} \, dt \nonumber \\
&=& \frac{1}{2}\left( 2f x + \frac{f^2 m}{\gamma^2} + T(x) \right).
\end{eqnarray} 
For the isothermal case, in the quasistatic limit where the velocity approaches zero ($v = 0$), $\dot{E}_{p} = 0$ and $\dot{H}_{d} = 0$. Far from the quasistatic limit, $\dot{E}_{p} = \dot{H}_{d} > 0$, which is expected as the particle operates irreversibly.

We now calculate other thermodynamic relations (in the absence of load) for different thermal arrangements. For a Brownian particle operating in a piecewise constant temperature:
\begin{equation} 
T(x) = \left\{
\begin{array}{cl}
T_{h}, & \text{for } 0 < x < L_0/2 \\
T_{c}, & \text{for } L_0/2 < x < L_0
\end{array}
\right.,
\end{equation}
while jumping from hot to cold heat bath,
\begin{eqnarray}
\Delta H_{d} = \frac{T_h - T_c}{2}.
\end{eqnarray}
as expected.

For a linearly decreasing temperature case,
\begin{eqnarray}
\frac{d H_{d}}{d x} = \frac{T_h - T_c}{2 L_o}. 
\end{eqnarray}
Integrating this equation, we obtain 
$
\Delta H_d = \int_{0}^{L_o} \frac{T_h - T_c}{2 L_o} \, dx = \frac{T_h - T_c}{2}
$.

For a quadratically decreasing temperature:
\begin{eqnarray}
\frac{d H_{d}}{d x} = \frac{(T_h - T_c) x}{L_o^2}.
\end{eqnarray}
Integrating this, we get 
$\Delta H_d = \int_{0}^{L_o} \frac{(T_h - T_c) x}{L_o^2} \, dx = \frac{T_h - T_c}{2}
$.
As expected  for the three types of thermal arrangements, $\Delta H_{d}={T_h-T_c \over 2}$. This also suggests  that $\Delta F(x)={T_c-T_h \over 2}$.This study indicates that the entropy production and energy extraction  increase as the temperature difference between the hot and cold reservoirs becomes larger. This increase  in entropy dissipation and entropy production is attributable to the greater thermal gradient driving the system further from equilibrium. Conversely, the change in free energy decreases as the temperature difference increases, showing  the  availability of free energy for performing work decreases when  higher thermal gradients is imposed.

These analysis collectively demonstrate that spatial variations in temperature inherently drive the system out of equilibrium. When accounting for the rate of heat dissipation due to particle recrossing at the interface of the heat reservoirs, the system exhibits irreversibility even in the quasistatic limit. Our findings further reveal that $H_{d}$ is significantly larger for a quadratically decreasing thermal gradient compared to a linearly decreasing one. Additionally, $H_{d}$ in the linearly decreasing scenario surpasses that in a piece-wise constant thermal configuration.

The main result of this study indicates that in the absence of load $f=0$,  rates such as  ${\dot e}_{p}$ and ${\dot h}_{d}$ go to zero.   However, this does not imply that   the system is approaching equilibrium.   In fact  our result shows that  the terms such as  entropy production and extraction  become non zero as long as a nonuniform  thermal arrangement is retained showing that  due to  heat exchange via kinetic energy,   the system is driven out of equilibrium.

\section{ More rigorous   approach for   linearly decreasing thermal  arrangement}  
	A similar set of assumptions is employed as in the derivation of Eq.~(22). Specifically, we consider the system in a steady state, assume a small particle mass \( m \), and impose periodic boundary conditions. Under these conditions, terms involving spatial derivatives in the Fokker–Planck equation are negligible compared to those in velocity space. The resulting distribution is obtained by solving the simplified steady-state equation (Eq. 2) and imposing normalization over both velocity and position, yielding an explicit expression for \( P(v, x) \) that is analytically tractable and physically consistent.
  After some algebra one gets  
\begin{widetext}
\begin{eqnarray}
P(x,v)&=& \nonumber \\
&&\frac{2 e^{\frac{m \left(f v-\frac{\gamma v^2}{2}\right)}{\gamma \left(T_h+\frac{(T_c-T_h)
x}{L_o}\right)}} \gamma^2 \sqrt{\frac{m}{T_c}} (T_c-T_h)}{L_o \sqrt{\pi } \left(\gamma^2 \left(-\frac{f^2
m}{\gamma^2 T_c}\right)^{3/2} T_c \text{Gamma}\left[-\frac{3}{2},-\frac{f^2 m}{2 \gamma^2 T_c}\right]+f^2
\sqrt{\frac{m}{T_c}} \sqrt{\frac{m}{T_h}} \sqrt{-\frac{f^2 m}{\gamma^2 T_h}} T_h \text{Gamma}\left[-\frac{3}{2},-\frac{f^2
m}{2 \gamma^2 T_h}\right]\right)}
\end{eqnarray}
\end{widetext}	
where Here, \(\Gamma(a, z)\) denotes the upper incomplete gamma function, defined by \(\Gamma(a, z) = \int_z^\infty t^{a-1} e^{-t} \, dt\).
 Here, the temperature decreases linearly from hot bath $T_h$ to cold bath $T_c$ along the reaction coordinate $L_0$.  We assume that viscous friction is independent of $x$. Please note that the probability distribution used in Section~III is based on a simplified steady-state solution that emphasizes analytical tractability over full normalization. This approximation allows us to investigate how entropy production and dissipation rates depend on system parameters without solving the full Fokker-Planck equation. In contrast, Section~IV adopts a more rigorous approach by explicitly imposing normalization of the joint distribution \( P(v, x) \) over both position and velocity. This refined treatment ensures consistency with the underlying stochastic dynamics and allows for exact evaluation of thermodynamic quantities across a wider parameter regime.

	Interestingly, in the absence of  a load and after some algebra, for a linearly decreasing temperature profile, we find that
	\begin{eqnarray}
{ dH_{d}\over dx}&=&{T_h-T_c \over 2 L_o}
\end{eqnarray}
This result is consistent with previous findings. Integrating this equation yields  $\Delta H_d=\int_{0}^{L_o} {T_h-T_c \over 2 L_o}dx={T_h-T_c \over 2}$. One can also   use the equipartition theorem to get 
	$
{H}_{d}={T(x) \over 2}
  $.

	We now integrate  the entropy production and extraction rates over $x$ and $t$. 
	Utilizing Eqs. (26), (12), and (13),  we derive the entropy production and extraction rates as:
	\begin{widetext}
\begin{eqnarray}
\dot{h}_d &=& \dot{e}_p \nonumber \\
&=& \frac{f^2 \sqrt{\frac{m}{T_c}} \left( 
\sqrt{2} L_0 \sqrt{\frac{m}{T_c}} \sqrt{-\frac{f^2 m}{\gamma^2 T_c}} T_c \, \Gamma\left(-\tfrac{1}{2}, -\frac{f^2 m}{2 \gamma^2 T_c} \right)
-
\sqrt{2} L_0 \sqrt{\frac{m}{T_h}} \sqrt{-\frac{f^2 m}{\gamma^2 T_h}} T_h \, \Gamma\left(-\tfrac{1}{2}, -\frac{f^2 m}{2 \gamma^2 T_h} \right)
\right)}
{2 \gamma L_0 \left( m T_c \, \mathrm{E}_{5/2}\left( \frac{f^2 m}{2 \gamma^2 T_c} \right)
-
\sqrt{\frac{m}{T_c}} \sqrt{\frac{m}{T_h}} T_h^2 \, \mathrm{E}_{5/2}\left( \frac{f^2 m}{2 \gamma^2 T_h} \right)
\right)}
\end{eqnarray}
\end{widetext}
where, \(\mathrm{E}_n(z)\) denotes the exponential integral of order \(n\), defined by
$\mathrm{E}_n(z) = \int_1^{\infty} \frac{e^{-z t}}{t^n} \, dt.$

 	In the limit $f \to 0$,  we find that ${\dot h}_{d}={\dot e}_{p}\to 0$ as anticipated.  Moreover,  in the limit $T_h \to T_c$,  we obtain ${\dot h}_{d}={\dot e}_{p}={L_0f^{2} \over \gamma T_c}$. 	The rates  ${\dot h}_{d}={\dot e}_{p}(x)$  depend  on the temperature and mass $m$. As the temperature difference increases, these rates also increase. Similarly, an increase in mass results in higher ${\dot h}_{d}$ and ${\dot e}_{p}$ values.

The average velocity does not depend on the mass and is found to be 
\begin{equation}  
 v={f\over \gamma },
 \end{equation} 
In the absence of an external force, the velocity approaches zero. However, this scenario does not imply that the system has reached equilibrium. Instead, it indicates that while the macroscopic average velocity $v$ is zero, the microscopic motion persists owing to thermal fluctuations. Consequently, the particle speed remains nonzero, allowing the continuous dissipation of heat via kinetic energy as the particle moves from the hotter  bath to the colder bath.   	 	The rate of heat dissipation is a critical factor in understanding the thermodynamics of a system. Utilizing Equations (14) and (15), we find that the rates converge to  
\begin{eqnarray}  {\dot H}_{d}&=&{\dot E}_{p}\nonumber \\   &=&{L_0f^2\over  \gamma}.    \end{eqnarray}
This result aligns with that obtained using Eq. (28). 	These thermodynamic relations do not depend on the temperature or mass $m$. 	  	Far from the temperature  gradient, the role of the heat exchange via kinetic energy  $H_{d}=-\left\langle {mv^2 \over 2}  +U(x)   \right\rangle$ can be  calculated in the absence of a ratchet potential as 
 \begin{eqnarray} {H}_{d}  &=& {r_1 +r_2 \over r_3}
 \end{eqnarray} 
where 
  \begin{widetext}
\begin{eqnarray}
r_1 &=& e^{\frac{f^2 m}{2 \gamma^2 T_c}} m \left( 
2 f^5 L_0 m^2 + 2 f^3 \gamma^2 L_0 m (T_c - 5 T_h) + 2 f \gamma^4 L_0 T_c (3 T_c - 5 T_h) + 6 f^4 m^2 (T_c - T_h) \right) \nonumber \\
&& + e^{\frac{f^2 m}{2 \gamma^2 T_c}} m \left( 
6 f^2 \gamma^2 m T_c (T_c - T_h) + 3 \gamma^4 T_c^2 (T_c - T_h) \right),
\\
r_2 &=& e^{\frac{f^2 m}{2 \gamma^2 T_h}} \sqrt{\frac{m}{T_c}} \sqrt{\frac{m}{T_h}} T_h \left( 
-2 f^5 L_0 m^2 + 8 f^3 \gamma^2 L_0 m T_h + 4 f \gamma^4 L_0 T_h^2 \right) \nonumber \\
&& + e^{\frac{f^2 m}{2 \gamma^2 T_h}} \sqrt{\frac{m}{T_c}} \sqrt{\frac{m}{T_h}} T_h \left( 
6 f^4 m^2 (T_h - T_c) + 6 f^2 \gamma^2 m T_h (T_h - T_c) + 3 \gamma^4 T_h^2 (T_h - T_c) \right),
\\
r_3 &=& 15 \gamma^4 (T_c - T_h) \Bigg( 
m T_c \, \mathrm{E}_{5/2} \left( \frac{f^2 m}{2 \gamma^2 T_c} \right) 
- \sqrt{\frac{m}{T_c}} \sqrt{\frac{m}{T_h}} T_h^2 \, \mathrm{E}_{5/2} \left( \frac{f^2 m}{2 \gamma^2 T_h} \right) 
\Bigg)
\end{eqnarray}
where $
\mathrm{E}_n(z) = \int_1^{\infty} \frac{e^{-z t}}{t^n} \, dt.
$

\end{widetext}

	Our  analysis indicates that ${H}_{d}$  increases  as mass $m$ and $T_h$  increase.  In the limit $T_h \to T_c$, we get: 
	\begin{eqnarray} 
	H_d&=&{1\over 2}\left(2fL_o+{f^2m \over \gamma^2}+T_c\right) 
	\end{eqnarray} which  agrees with Eq. (31). In the limit $f \to 0$, Eq. (44) converges to  
	\begin{eqnarray}  { H}_{d}&=&\frac{3 T_c^2 \sqrt{\frac{1}{T_h}}-3 \sqrt{\frac{1}{T_c}} T_h^2}{10 T_c \sqrt{\frac{1}{T_h}}-10 \sqrt{\frac{1}{T_c}} T_h}.    
	\end{eqnarray} 	In the limit $T_h \to T_c$, the above equations reduces to   
	\begin{eqnarray} 
	{H}_{d} &=&{T_c\over 2}.  
		\end{eqnarray} 
			Further analysis  indicates that this  kinetic  energy is equally partitioned  along length $L_o$ and the  change in kinetic energy is zero across $L_o$ for the isothermal case.  The situation differs for the non-isothermal case because  $ \Delta H_{d} >0$ shows that even in the absence of a load, the system is driven out of equilibrium owing to the thermal gradients.

In this section,  we  rigorously  solve several thermodynamic relations  for a system  with linearly decreasing thermal gradient. By solving the probability distribution function, we derive analytical expressions for the entropy production and extraction rates. We find that the steady-state probability distribution depends on the position, velocity,  and temperature of the hot and cold baths.  The energy dissipation rate strictly depends on  parameters such as force, friction, and  length. At steady state,   the entropy production and heat dissipation rates are  found to be equivalent. The average velocity of the system is proportional to the applied force and viscous friction.  Moreover, our findings demonstrate that the average  kinetic energy   depends strictly  on   the strength of    the noise  intensity. We believe that  because our work offers insight   into optimizing energy use and the role of thermal arrangements,  this study helps to understand  the physics of  nanoscale heat engines, biological systems, and microscale robotics.   The results discussed in this section also  provide a better understanding  of  non-equilibrium thermodynamics and energy partitioning in temperature-gradient-driven systems.  Furthermore, our analysis reveals that thermal gradients alone can drive systems out of equilibrium.    In future research, it is vital to  explore the effects of nonlinear thermal gradients and external potentials so that we can obtain a complete   understanding of nonequilibrium transport features  in complex systems.

A key theoretical result emerging from our analysis is that both the entropy production rate, \(\dot{e}_p\), and the entropy extraction rate, \(\dot{h}_d\), approach zero in the limit of the vanishing external force, even in the presence of a nonuniform spatial temperature profile. As derived in Section III (see Eqs.~(24), (27), and (41)), these rates exhibit a quadratic dependence on the external load \(f\), namely \(\dot{e}_p \sim \dot{h}_d \sim f^2 / \gamma T(x)\). Consequently, when \(f \to 0\), the local entropy production and extraction rates vanish throughout the system, independent of the specific thermal configuration, be it linear, quadratic, or piecewise constant.  However, this asymptotic decay of the instantaneous rates does not imply that the system has reached thermodynamic equilibrium.  On the contrary this indicates that  the presence of spatially varying temperature leads to irreversible dynamics driven by stochastic fluctuations and heat transfer via kinetic energy. This is substantiated by the fact that the total entropy production and extraction, obtained through integration over time or space, remain strictly positive, as shown analytically in Eqs.~(31)--(33), (48)--(50), and (65).  These integrated quantities capture the cumulative effect of microscopic energy exchange across the thermal gradient.  Therefore, a critical distinction arises between rate-based and integrated measures of irreversibility while the former may vanish in the zero-force limit, the latter retains finite values, reflecting the sustained nonequilibrium character of the steady state.  This observation emphasizes that vanishing entropy production or extraction rates should not be interpreted as indicative of equilibrium, particularly in underdamped systems with spatial thermal asymmetry.  Instead, the system remains thermodynamically irreversible due to the exchange of  kinetic energy across the temperature landscape.

To avoid ambiguity, we now distinguish clearly between instantaneous entropy production rates and cumulative (or total) entropy production. The rate $\dot{e}_p(t)$ characterizes the local, instantaneous entropy generated at time $t$, whereas the total entropy production $\Delta e_p(t)$ is defined by the time integral $\Delta e_p(t) = \int_0^t \dot{e}_p(t')\,dt'$. Importantly, it is possible for $\dot{e}_p \to 0$ as $f \to 0$, while $\Delta e_p > 0$ remains finite due to prior nonequilibrium dynamics. A vanishing entropy production rate does not necessarily imply equilibrium. The system can be out of equilibrium due to spatial asymmetries, thermal gradients, or kinetic energy exchange while   $\dot{e}_p = 0$ at a given time.

\section{ Ratchet potential  coupled with piecewise thermal arrangement}  

  In this section, we  consider   a Brownian particle  that  walks  along  a  ratchet potential  
	\begin{equation} U_{s}(x)=\left\{ \begin{array}{ll} 2U_{0}\left({x\over L_{0}}\right),& \text{if}~~~ 0 \le x \le {L_{0}\over 2}; \\ 2U_{0}\left(1-{x\over L_{0}}\right),& \text{if} ~~~{L_{0}\over 2} \le x \le L_{0} \end{array} \right. 
	\end{equation} where   $U(x)=U_{s}(x)+fx$.  Here, the potential  $U_{s}(x)$ denotes a ratchet potential  and  $f$ is the load.  $U_0$ and $L_0$ denote   the height of the barrier and width of the ratchet potential, respectively. The ratchet  potential is coupled with  a  spatially varying temperature    
\begin{equation}
T(x)=\left\{
\begin{array}{ll}
T_{h},& \text{if} ~~~0 \le x \le {L_{0}\over 2};\\
T_{c},& \text{if} ~~~ {L_{0}\over 2} \le x \le L_{0}
\end{array}
\right.
\end{equation}
Moreover, the potential $U_s(x)$ and temperature $T(x)$ are assumed to be periodic with a period of $L_0$, such that $U_s(x+L_0)=U_s(x)$ and $T(x+L_0)=T(x)$.

Assuming the probability distribution is normalized with respect to both position and velocity,   we solve  Eq. (2) as
\begin{widetext}
\begin{eqnarray}
P(x,v)&=& \nonumber \\
&&e^{\frac{m \left(f L_o v+2 U_0 v+\frac{1}{2} \gamma L_o v^2-4 U_0 v \Theta\left[-\frac{L_o}{2}+x\right]\right)}{\gamma
L_o \left(-T_h+(-T_c+T_h) \Theta\left[-\frac{L_o}{2}+x\right]\right)}}/\left(\frac{L_o \sqrt{\frac{\pi
}{2}} \left(e^{\frac{m (f L_o-2 U_0)^2}{2 \gamma^2 L_o^2 T_c}} \sqrt{\frac{m}{T_c}} T_c+e^{\frac{m
(f L_o+2 U_0)^2}{2 \gamma^2 L_o^2 T_h}} \sqrt{\frac{m}{T_h}} T_h\right)}{m}\right).
\end{eqnarray}
\end{widetext}	
Here, $\Theta\left[x\right]$ denotes the Heaviside function.

The average velocity depends on the mass and  is given by
\begin{widetext}
\begin{eqnarray}
v&=& \nonumber \\
&&-\frac{e^{\frac{m (f L_o-2 U_0)^2}{2 \gamma^2 L_o^2 T_h}} \sqrt{\frac{m}{T_c}}
(f L_o-2 U_0)+e^{\frac{m (f L_o+2 U_0)^2}{2 \gamma^2 L_o^2 T_c}} \sqrt{\frac{m}{T_h}} (f L_o+2
U_0)}{\gamma L_o \left(e^{\frac{m (f L_o-2 U_0)^2}{2 \gamma^2 L_o^2 T_h}} \sqrt{\frac{m}{T_c}}+e^{\frac{m
(f L_o+2 U_0)^2}{2 \gamma^2 L_o^2 T_c}} \sqrt{\frac{m}{T_h}}\right)}.
\end{eqnarray}
\end{widetext}
In the limit $U_0 \to 0$, we obtain $v = \frac{f}{\gamma}$. However, in the limit $f \to 0$, we obtain
\begin{widetext}
\begin{eqnarray}
v&=& \nonumber \\
&&-\frac{2 e^{\frac{2 m U_0^2}{\gamma^2 L_o^2 T_h}} \sqrt{\frac{m}{T_c}} U_0-2 e^{\frac{2 m
U_0^2}{\gamma^2 L_o^2 T_c}} \sqrt{\frac{m}{T_h}} U_0}{e^{\frac{2 m U_0^2}{\gamma^2 L_o^2 T_h}} \gamma
L_o \sqrt{\frac{m}{T_c}}+e^{\frac{2 m U_0^2}{\gamma^2 L_o^2 T_c}} \gamma L_o \sqrt{\frac{m}{T_h}}},
\end{eqnarray}
\end{widetext}

Whenever we plot  figures, we use  the following   rescaled mass  $\tilde{m} = m / (\gamma \kappa)$  where $\kappa = \gamma L^2 / (k_B T)$ is the characteristic time scale. The load and the barrier height  are also rescaled as $\tilde{f} = f L / (k_B T)$ and $\tilde{U}_0 = U_0 / (k_B T)$  while the length and temperature  are also rescaled as 
 $\tilde{x} = x / L$  and $\tau={T_h\over T_c}$.
  From now on, all variables are considered dimensionless, and the tildes are omitted for simplicity.
\begin{figure}[ht]
\centering
%\subfigure[Bild a.] % caption for subfigure a
{
    %\label{fig:sub:b}
    \includegraphics[width=8cm]{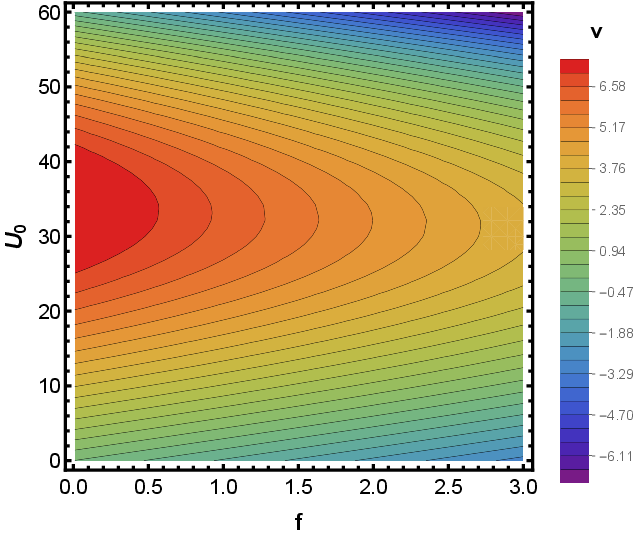}
}
\caption{ (Color online) Contour plot of the steady-state velocity $v$ of a Brownian particle in an underdamped medium as a function of the external load $f$ and potential barrier height $U_0$. for fixed $m=0.0001$, $L=1$ and $\tau=2$. } 
\label{fig:sub} % caption for the whole figure
\end{figure}
 Via Eq. (54), it is evident  that the particle experiences a unidirectional current in the presence of a load $f$ or non-uniform temperature. Notably, when $T_h>T_c$ and load $f$ is small,  the particle velocity  becomes positive. Consequently, the particles operate   as heat engines. However, as the load increases, the velocity decreases. For a sufficiently large load (see Fig. 1), the particle  current reverses direction, causing the motor to function as a refrigerator. This is consistent with the results of previous studies \cite{mu17,muu17}.  The same figure  also shows that the current  monotonously increases,  and at a larger  barrier
 height, it starts  to  decrease,   which is in agreement with the results of previous studies \cite{mu17,muu17}.

\begin{figure}[ht]
\centering
%\subfigure[Bild a.] % caption for subfigure a

%\subfigure[Bild b.] % caption for subfigure b
{
    %\label{fig:sub:b}
    \includegraphics[width=8cm]{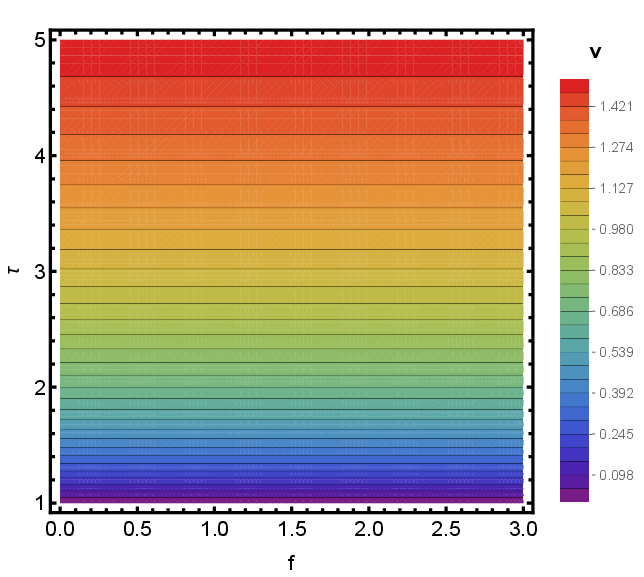}
}
\caption{ (Color online) Contour plot of the steady-state velocity $v$ of a Brownian particle in an underdamped medium as a function of the external load $f$ and potential barrier height $\tau$  for fixed $U_0=2.0$,  $L=1$ and $m=0.001$. } 
\label{fig:sub} % caption for the whole figure
\end{figure}
 Exploiting Eq. (54), one can also see that   the particle also exhibits unidirectional current in the absence of a load, as long as  a distinct temperature difference is retained  between the hot and cold baths, as depicted in Fig. 2.   Our  analysis also  indicates that   the velocity decreases as the mass increases,  suggesting the possibility of separating particles based on their mass. Moreover, the velocity increases as the temperature of the hot bath increases.  In other words,  not only the load but also the mass $m$ is responsible   for the current reversal.  At larger masses, current reversal can  occur, depending on  the magnitude of  the other parameters.

It is natural to interpret the directed motion observed in our model, particularly  in the absence of an external load, as a manifestation of thermophoretic drift. In this case,  a Brownian particle experiences net transport due to spatial variations in temperature. This type of motion, often referred to as noise-induced or thermally driven transport  and it is  a well-studied phenomenon in soft-matter systems. Our results provide a concrete realization of this mechanism within the underdamped regime, where we derive explicit steady-state velocity distributions and average drift expressions that depend on particle mass and spatial inhomogeneities. Moreover, the results obtained in this work conceptually align with the rigorous framework developed by Hottovy et al.~\cite{c8}, who analyzed the Smoluchowski–Kramers limit of stochastic differential equations with general, state-dependent friction and noise coefficients. They demonstrated that the resulting overdamped dynamics exhibit a noise-induced drift term, which emerges from the interplay of inertia and spatial inhomogeneity—corrections that are absent in naive overdamped approximations. While our approach does not rely on taking the small-mass limit, but instead remains within the underdamped regime, the qualitative behavior is consistent with their theory.

 \begin{figure}[ht]
\centering
%\subfigure[Bild a.] % caption for subfigure a

%\subfigure[Bild b.] % caption for subfigure b
{
    %\label{fig:sub:b}
    \includegraphics[width=5cm]{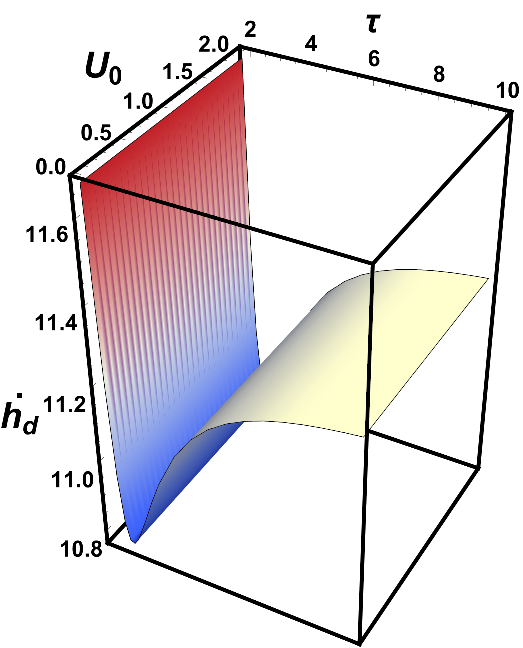}
}
\caption{ (Color online) Plot of ${\dot h}_{d}={\dot e}_{p}$ as a function of  the rescaled temperature $\tau$ and mass $U_0$   for fixed $f=0.6$, $m=1.0$ and   $L=1$. } 
\label{fig:sub} % caption for the whole figure
\end{figure}
The  plot of ${\dot h}_{d}={\dot e}_{p}$ as a function of  rescaled temperature $\tau$ and mass $U_0$ indicates that the entropy production or  extraction rates  decreases with $\tau$ and  as $\tau$ further increases,  these rates  increase with $\tau$.

 In summary,   even in the absence of an external force, the particle exhibits unidirectional motion as long as a distinct temperature difference was retained across the reaction coordinate.  Both the asymmetry in  the thermal distribution as well the external load dictate  particle motion.  Model ingredients such as barrier height, external force,  and noise intensity  affect the magnitude of the velocity.  The direction of the velocity is dictated solely by the external load, where the motor acts as a heat engine or refrigerator depending on the magnitude of the applied load.  These results have profound applications in microscale and nanoscale transport systems, where controlled particle movement is essential. The fact that the motor can exhibit directed motion even in the absence of a load shows that the velocity of the motor can be manipulated by varying the magnitude of the noise intensity.  The velocity of the particles strictly depends on their mass, indicating that  particles with different masses can be sorted based on their physical properties.

  In terms of implications, this study  stresses how  optimizing system parameters, such as barrier height and temperature differences, is  vital for  achieving  the desired particle velocity.   The results  also provide insights into  the behavior of tiny molecular motors as the model parameters   are varied.  This work also  has a direct impact on understanding the energy-harvesting process at microscopic or nanoscopic scales and provides valuable insights into nonequilibrium statistical mechanics and stochastic thermodynamics. As a future  direction,  this study paves the way for exploring the influence of additional factors, such as particle interactions and complex potential landscapes, to further refine the understanding of Brownian ratchet systems.

{\it Other thermodynamic relations\textemdash}
Similarly,  the entropy production and extraction rates can be obtained  by employing  Equations (26), (12), and (13).  After some algebra  one gets
	\begin{widetext}
 \begin{eqnarray} 
{\dot h}_{d}&=&\frac{\sqrt{m} \left(\frac{e^{\frac{m (f L_o-2 U_0)^2}{2 \gamma^2 L_o^2 T_h}} (f L_o-2
U_0)^2}{\sqrt{T_h}}+\frac{e^{\frac{m (f L_o+2 U_0)^2}{2 \gamma^2 L_o^2 T_c}} (f L_o+2 U_0)^2}{\sqrt{T_c}}\right)}{\gamma
L_o^2 \left(e^{\frac{m (f L_o+2 U_0)^2}{2 \gamma^2 L_o^2 T_c}} \sqrt{m T_c}+e^{\frac{m (f
L_o-2 U_0)^2}{2 \gamma^2 L_o^2 T_h}} \sqrt{m T_h}\right)}.
   \end{eqnarray}
	\end{widetext}
 In the limit $T_h \to T_c$,  we obtain ${\dot h}_{d}={\dot e}_{p}=L_0{f^{2} \over \gamma T_c}$. 	The values of ${\dot h}_{d}={\dot e}_{p}$ are directly influenced by the temperature and mass $m$. As the temperature  of  the hot bath  increases, these rates also increase. In addition, a higher mass leads to greater values of ${\dot h}_{d}$ and ${\dot e}_{p}$.

 	The term related to the heat dissipation rate can be obtained using Eqs.   (14) and (15), respectively: One gets  
		\begin{widetext}
 \begin{eqnarray} 
{\dot H}_{d}
 &=&\frac{\sqrt{m} \left(e^{\frac{m (f L_o-2 U_0)^2}{2 \gamma^2 L_o^2 T_h}} \sqrt{T_h}
(f L_o-2 U_0)^2+e^{\frac{m (f L_o+2 U_0)^2}{2 \gamma^2 L_o^2 T_c}} \sqrt{T_c} (f L_o+2 U_0)^2\right)}{\gamma
L_o^2 \left(e^{\frac{m (f L_o+2 U_0)^2}{2 \gamma^2 L_o^2 T_c}} \sqrt{m T_c}+e^{\frac{m (f
L_o-2 U_0)^2}{2 \gamma^2 L_o^2 T_h}} \sqrt{m T_h}\right)}
   \end{eqnarray}
		\end{widetext}
As $f$ approaches zero, ${\dot H}_{d}$ converges to zero as expected. In contrast, when $T_h$ approaches $T_c$, we obtain that $\dot{H}_{d}\to ={L_0f^{2} \over \gamma}$.

Following the  approach discussed above,  by intergarting out  time only,  after some algebra  one cam also show that  	\begin{eqnarray} { dH_{d}\over dx}={T_h-T_c \over 2 } \end{eqnarray} This result is in agreement with the previous findings.

Furthermore,   the heat exchange via  kinetic energy ($H_{d}=-\left\langle {mv^2 \over 2}  + U(x)  \right\rangle$) can be determined as 
\begin{eqnarray}
{H}_{d}
&=& {r_1 +r_2 \over r_3}
\end{eqnarray}
where 
 \begin{widetext}
\begin{eqnarray}
r_1 &=& \sqrt{m} \, e^{\frac{m (f L_0 - 2 U_0)^2}{2 \gamma^2 L_0^2 T_h}} \sqrt{T_h} 
\left( L_0^2 \left( 2 f^2 m + \gamma^2 (f L_0 + 2 T_h) \right) - 8 f L_0 m U_0 + 8 m U_0^2 \right), \\
r_2 &=& \sqrt{m} \, e^{\frac{m (f L_0 + 2 U_0)^2}{2 \gamma^2 L_0^2 T_c}} \sqrt{T_c} 
\left( L_0^2 \left( 2 f^2 m + \gamma^2 (3 f L_0 + 2 T_c) \right) + 8 f L_0 m U_0 + 8 m U_0^2 \right), \\
r_3 &=& 4 \gamma^2 L_0^2 \left( 
e^{\frac{m (f L_0 + 2 U_0)^2}{2 \gamma^2 L_0^2 T_c}} \sqrt{m T_c} 
+ e^{\frac{m (f L_0 - 2 U_0)^2}{2 \gamma^2 L_0^2 T_h}} \sqrt{m T_h} 
\right).
\end{eqnarray}
\end{widetext}

According to our analysis, ${H}_{d}$  increases as mass $m$ and $T_h$ increase. In the limit $T_h \to T_c$, we obtain
\begin{eqnarray}
H_d={1\over 2}\left(2f L_0+{f^2m \over \gamma^2}+T_c\right)
\end{eqnarray}
which agrees with Eq. (31). In the limit $f \to 0$, Eq. (59) converges to
 \begin{eqnarray} 
{ H}_{d} &=&
\frac{1}{2} \left( T_c + T_h - \sqrt{T_c T_h} \right).
   \end{eqnarray}
	In the limit $T_h \to T_c$, the above equations reduces to  
\begin{eqnarray} 
{H}_{d}&=&{T_c\over 2}.
   \end{eqnarray}
	Further analysis suggests that the kinetic energy is equally partitioned along length $L_o$, resulting in a zero change in kinetic energy across $L_o$ for the isothermal case. In contrast, the non-isothermal case  $\Delta H_{d} > 0$ indicates that the system is driven out of equilibrium due  to thermal gradients, even in the absence of a load.
	\begin{figure}[ht]
\centering
%\subfigure[Bild a.] % caption for subfigure a

%\subfigure[Bild b.] % caption for subfigure b
{
    %\label{fig:sub:b}
    \includegraphics[width=5cm]{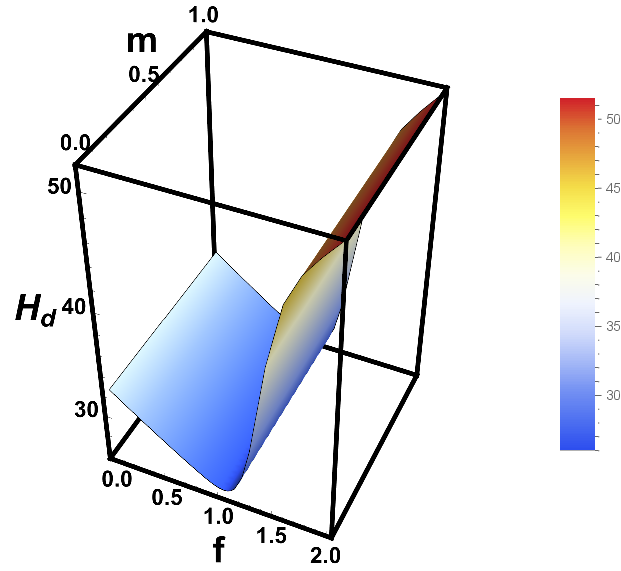}
}
\caption{ (Color online) Plot of $H_d$ as a function of  the force $f$  and mass $m$   for fixed $U_0=4.0$, $\tau=2.0$ and   $L=1$. } 
\label{fig:sub} % caption for the whole figure
\end{figure}
 Figure 4 shows a plot of $H_d$ as a function of  force $f$  and mass $m$   for fixed $U_0=4.0$, $\tau=2.0$ and   $L=1$.  The plot   shows that  $H_d$   decreases with the load and converges to the minimum value.  When the force was increased further, $H_d$  increased.
\begin{figure}[ht]
\centering
%\subfigure[Bild a.] % caption for subfigure a

%\subfigure[Bild b.] % caption for subfigure b
{
    %\label{fig:sub:b}
    \includegraphics[width=5cm]{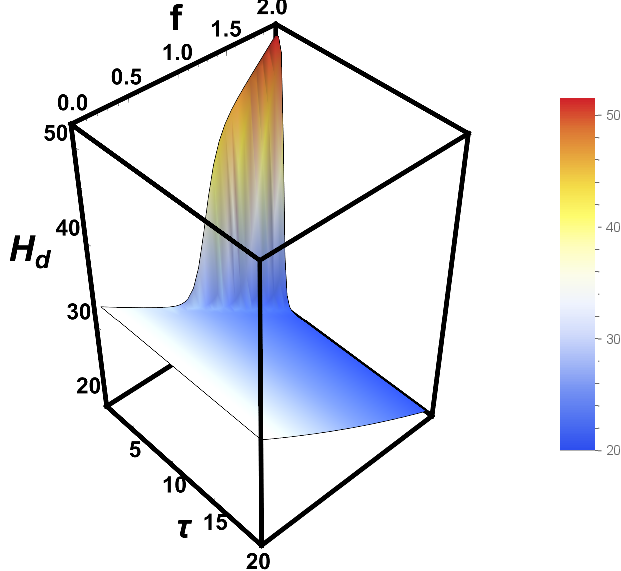}
} 
\caption{ (Color online)  Plot of $H_d$ as a function of  the force $f$  and mass $\tau$   for fixed $U_0=4.0$, $m=1.0$ and   $L=1$. } 
\label{fig:sub} % caption for the whole figure
\end{figure}
In Fig. 5, we plot $H_d$ as a function of  force $f$  and mass $\tau$   for fixed $U_0=4.0$, $m=1.0$ and   $L=1$. The figure shows that  as the temperature difference between  the hot and cold baths increases, $H_d$ increases.

 Entropy production and extraction rates can be derived using established thermodynamic equations, which show a direct dependence on temperature and mass. As the temperature of the hot bath increases, the rates also increase. In particular,  an increase in mass  results in greater entropy production. Additionally, the heat dissipation rate follows a similar pattern; as a result, it approaches zero as the external load approaches zero.

 Furthermore, the    kinetic energy is partitioned equally along   the system length, and the change in  total  kinetic energy depends on the magnitude of the noise  intensity. For an isothermal system, the kinetic energy remains constant, whereas the non-isothermal conditions   deviate from zero, indicating  that the system is driven out of   equilibrium. This analysis confirms that thermal asymmetry alone can drive transport out of equilibrium,   which in turn   provides  insights for optimizing energy-efficient microscale devices.

\section{The role of   multiplicative noise and   viscous friction}

\subsection{ Multiplicative noise for underdamped case}

Now, rather than focusing on additive noise,   let us    explore the multiplicative noise of thermodynamic features, where the noise amplitude varies spatially \cite{mar12}.  In a recent study that  considered multiplicative noise,  the noise-induced phase separation behavior was explored  \cite{mar10}.       In this work, we  study  how the entropy, entropy production, and extraction rate depend  on the strength  of the background noise by  solving the model by considering the temperature-dependent noise   $T (x) = \sqrt{D}|x|^{-z/2}$ in the absence of any external potential.  After some algebra, the    entropy production and extraction rate  are calculated  as 
\begin{eqnarray}  
\dot{h}_{d}(x) &=& \dot{e}_{p}(x) \\
 &=& \frac{f^{2}}{\gamma \sqrt{D}|x|^{-z/2}}
\end{eqnarray}
The  above equation clearly  shows that   the entropy production  or extraction  rates are significantly high  in regions where $( |x| )$ is smaller.  These rates are  considerably small   in regions where \( |x| \) is larger. The overall result shows that a stronger noise intensity  leads to localized energy dissipation, which in turn influences the overall efficiency of energy utilization within the system.

The total entropy extraction  rate is given by
\begin{eqnarray}  
\dot{H}_{d}(x) &=& \frac{f^2}{\gamma}.
\end{eqnarray}
This suggests that the total entropy production is independent of the temperature, and only the external force  affects the dynamics. As the force increases,  the term related to the entropy production increases. On the other hand, the  entropy extraction is given by
\begin{eqnarray}  
H_{d}(x) &=& \frac{1}{2}\left(2fx + \frac{f^2m}{\gamma^2} + \sqrt{D}|x|^{-z/2}\right)
\end{eqnarray}
which shows that  entropy extraction is an increasing function of the   deterministic driving forces and stochastic noise strength $D$.  This clearly shows that the interplay between the deterministic and random noise terms  dictates the  underdamped dynamics.
In the absence of an applied load, the entropy extraction reduces to 
\begin{eqnarray}  
H_{d}(x) &=& \frac{\sqrt{D}|x|^{-z/2}}{2},
\end{eqnarray}
for $z \leq 2$. This suggests that, in the absence of external forces, noise solely governs entropy dynamics.  This  effect becomes  dominant for smaller values of \( |x| \). This behavior is essential for understanding thermal fluctuations in nanoscale systems, where noise-induced effects significantly affect the stability and performance.

The  analysis shows that entropy production depends on the square of the applied force and is inversely related to the noise amplitude. Higher noise levels, represented by larger values of \( D \) and \( |x|^{-z/2} \), lead to lower entropy production, indicating an improved energy dissipation efficiency. This understanding has significant applications across various fields, such as biological systems, where stochastic forces play a crucial role in intracellular transport, influencing the movement of molecules and organelles within cells, and nanotechnology, where noise analysis is essential for designing and optimizing nanoscale devices. Note that  thermal fluctuations at such small scales can affect the stability and performance. Climate modeling also benefits from these insights because spatially varying noise significantly affects atmospheric and oceanic systems, leading to more accurate predictions of climate patterns and variability. In robotics and autonomous systems, understanding noise behavior enhances particle motion  and adaptability, allowing robots operating in uncertain environments to leverage noise analysis to improve responsiveness and efficiency. Energy harvesting systems can optimize efficiency by better managing noise interactions and harnessing stochastic fluctuations to maximize energy conversion and performance. In addition, medical diagnostics can benefit from noise analysis by improving the interpretation of biological signals, leading to enhanced diagnostic tools and treatment strategies that account for physiological noise effects.

\subsection{ The role of  viscous friction}

 Several   experimental  works  show that the viscous friction $\gamma$ depends   on temperature, and  it  decreases  exponentially as the temperature   increases  ($\gamma=Be^{-A T}$),  as  first proposed by Reynolds \cite{mar15}.  In our previous study, we explored   the  dependence of   several thermodynamic relations considering  viscous friction $\gamma (T(x))$ which  decreases exponentially as temperature $T(x)$ increases.  However, we did not explore how   thermodynamic  relations, such as entropy production and entropy exchange rates, behave. Thus, in this work, we will address this issue.

 Let us consider    viscous  friction $\gamma$ that has an exponential temperature dependence as  
 \begin{eqnarray}  \gamma (T) & =&Be^{-A T(x)}. \end{eqnarray} 
Here, $A$ and $B$ are constants that characterize the system. Here 
 \begin{equation}    T(x)=T_{h}+{x(T_{c}-T_{h})\over L_o}    \end{equation} 
decreases  linearly.

Imposing   the requirement that the probability distribution be normalized to unity for all the positions, after some algebra we get  
\begin{eqnarray} 
{\dot h}_{d}(x)&=&{\dot e}_{p}(x) \nonumber \\
&=&  \frac{e^{A \left( T_h + (T_c - T_h)x \right)} f^2}{B \left( T_h + \frac{(T_c - T_h)x}{L_0} \right)}.
   \end{eqnarray}
	This shows that entropy production increases with temperature, where higher temperatures lead to greater energy dissipation. The temperature gradient highlights the spatial dependence of dissipation, which is important for understanding energy transport in systems with varying thermal conditions.

The term related to the entropy extraction rate is given as 	
	\begin{eqnarray} 
{\dot H}_{d}(x)
&=&\frac{e^{A \left( T_h + (T_c - T_h)x \right)} f^2}{B} 
   \end{eqnarray}
	for the linearly decreasing temperature case.  This  indicates that the total dissipation remains constant across the system, implying a uniform force-driven energy loss.

	On the other hand, one gets   
	\begin{eqnarray} 
{H}_{d}(x)
&=&-\frac{1}{2} \left( \frac{e^{2 A \left( T_h + (T_c - T_h)x \right)} f^2}{B^2} + T_h + 
2 f x + \frac{(T_c - T_h)x}{L_0} \right)
   \end{eqnarray}
for  the linearly decreasing  case.  This  equation highlights the combined effects of deterministic forces and noise, demonstrating how entropy evolves with the temperature and position.

Note that  viscous friction  does  affect $H_d$. As discussed before, 	for  a linearly decreasing temperature in the absence of a load, ${ dH_{d}\over dx}={T_h-T_c \over 2 L_o}$. 	 Furthermore, this study also shows that when the inhomogeneity in viscous friction increases (i.e., when the parameter $A$ increases), the rates of entropy extraction or production correspondingly increase. This indicates that the  change  in viscous friction plays a significant role in enhancing the non-equilibrium processes within the system.

Considering a temperature-dependent viscous friction model is vital because such an approach provides an accurate representation of particle dynamics in biological systems. Most  traditional models  assume constant vicious friction.  However,  in most biological systems, thermal fluctuation significantly affects the viscosity and the resulting dynamics, such as particle movement. For instance,  the temperature  in the cytoplasm directly influences the viscosity of the fluid within the cytoplasm.  This, in turn, affects essential processes such as intracellular transport, protein folding, and signal transduction. Particularly in the exponential decay model, the viscous friction decreases as the temperature increases.  As the temperature increases, the  viscous friction strength decreases owing to  thermal agitation, as confirmed by the experiment.  Thus, this approach provides a deeper understanding of biological transport mechanisms,  such as   calcium signaling in cardiac tissue and the motion of molecular motors in a cell. 

Here, we want to add our perspective regarding why studying entropy  production is vital  in biological systems where the viscosity in these systems is temperature-dependent.  Entropy often   acts as a measure of energy dissipation and system irreversibility. Entropy  serves as a measure of energy and transport efficiency.  Because thermal fluctuations within biological systems affect viscosity and particle motion,  the temperature-dependent friction model enables us to study how the model ingredients affect  the dynamics of the system. Moreover,  monitoring entropy change not only helps optimize energy consumption but also offers valuable diagnostic insights. High entropy production may be related to  increased dissipation and system instability,  whereas  lower entropy suggests more efficient transport and energy utilization. Thus,  in order to have a predictive model that helps to optimize transport mechanisms while minimizing dissipation,   exploring the system entropy is vital.  Despite the potential of the model to enhance predictive capabilities, challenges remain in accurately characterizing parameters and addressing the complexity of biological environments. However, advancements in imaging and particle-tracking technologies continue to refine these models, unlocking new possibilities in biomedical research and innovation.

\section{Summary and Conclusion}

In this study, we  explore the thermodynamic  features  of a single Brownian particle that moves  in an underdamped medium subjected  to  quadratically decreasing, linearly decreasing, and piecewise constant-temperature distributions along the reaction coordinates. Using  a rigorous analytical approach, we  present  general thermodynamic relations that provide deeper insight into how temperature gradients govern energy exchange and irreversibility in nonequilibrium systems.  

A central question addressed  in this study  includes   whether vanishing entropy production or extraction rates necessarily indicate equilibrium. The results of this work exhibit that even in the absence of an external force, entropy production or extraction rates  can approach zero, while total entropy production and extraction remain finite. This  clearly indicates  that the system maintains a nonequilibrium steady state due to heat transfer via kinetic energy. We  also examine the influence of spatially varying viscous friction, showing  that although most thermodynamic rates decay over time, irreversible heat  flow persists owing to continuous heat exchange across the temperature gradient.

A central conclusion of this study is that both the entropy production rate, \(\dot{e}_p\), and the entropy extraction rate, \(\dot{h}_d\), vanish in the limit of zero external force, even when a spatially varying temperature profile is present. As shown in Eqs.~(24), (27), and (41), these rates scale as \(f^2 / \gamma T(x)\), and thus asymptotically decay as \(f \to 0\). However, this vanishing of instantaneous rates does not imply equilibrium. The system remains out of equilibrium due to persistent irreversible heat flow via kinetic energy, sustained by the thermal gradient. This is evident from the finite total entropy production and extraction, \(\Delta e_p = \int \dot{e}_p \, dt\) and \(\Delta h_d = \int \dot{h}_d \, dt\), as demonstrated in Eqs.~(31)--(33), (48)--(50), and (65). These findings underscore that a vanishing rate does not equate to thermodynamic equilibrium, but rather reflects a nonequilibrium steady state driven by continuous stochastic heat exchange.

Furthermore, we show that  a Brownian particle in a ratchet potential with spatially varying temperature exhibits directed motion even without an external load. This transport emerges from thermal asymmetry and can be controlled by tuning parameters such as particle mass, barrier height, and noise intensity. We believe  that  the results  presented in this work have practical implications  that include  particle separation, microfluidic transport, and sensor design in  microscale and nanoscale systems. This work also  provides a robust theoretical framework for understanding the thermally driven motion  of Brownian motors  driven by temperature asymmetry.

	\section*{Acknowledgment}
I would like to thank  Mulu  Zebene and Asfaw Taye for the
constant encouragement.


\begin{thebibliography}{80}

\bibitem{mu1} H. Ge and H. Qian, Phys. Rev. E {\bf 81}, 051133 (2010).
\bibitem{mu2} T. Tome and M. J. de Oliveira, Phys. Rev. Lett. {\bf 108}, 020601 (2012).
\bibitem{mu3} J. Schnakenberg, Rev. Mod. Phys. {\bf 48}, 571 (1976).
\bibitem{mu4} T. Tome and M.J. de Oliveira, Phys. Rev. E {\bf 82}, 021120 (2010).
\bibitem{mu5} R.K.P. Zia and B. Schmittmann, J. Stat. Mech. {\bf P07012} (2007). 
\bibitem{mg11} M. A. Taye,  Phys. Rev. E {\bf 103},  042132 (2021). 
\bibitem{mu6} U. Seifert, Phys. Rev. Lett. {\bf 95}, 040602 (2005). 
\bibitem{mu7} T. Tome, Braz. J. Phys. {\bf 36}, 1285 (2006).
\bibitem{mu8} G. Szabo, T. Tome  and I. Borsos, Phys. Rev. E {\bf 82}, 011105 (2010).
\bibitem{mu9} B. Gaveau, M. Moreau and L.S. Schulman, Phys. Rev. E {\bf 79}, 010102 (2009).
\bibitem{mu10} J.L. Lebowitz and H. Spohn, J. Stat. Phys. {\bf 95}, 333 (1999).
\bibitem{mu11} D. Andrieux and P. Gaspar, J. Stat. Phys. {\bf 127}, 107 (2007). 
\bibitem{mu12} R.J. Harris and G.M. Schutz, J. Stat. Mech.  {\bf P07020} (2007). 
\bibitem{ta1}T.~Tomé and M.~J.~de~Oliveira, Phys.\ Rev.\ E {\bf 91}, 042140 (2015).
\bibitem{mu13} J.-L. Luo, C. Van den Broeck, and G. Nicolis, Z. Phys. B {\bf 56}, 165 (1984). 
\bibitem{mu14} C.Y. Mou, J.-L. Luo, and G. Nicolis, J. Chem. Phys. {\bf 84}, 7011 (1986). 
\bibitem{mu15} C. Maes and K. Netocny, J. Stat. Phys. {\bf 110}, 269 (2003).
\bibitem{mu16} L. Crochik and T. Tome, Phys. Rev. E  {\bf 72}, 057103 (2005). 
\bibitem{mu17} M. Asfaw,  Phys. Rev. E {\bf 89}, 012143 (2014).
\bibitem{muu17} M. Asfaw,  Phys. Rev. E {\bf 92},  032126 (2015).
\bibitem{mu25} K. Brandner, M. Bauer, M. Schmid  and U. Seifert,  New. J. Phys. {\bf 17}, 065006 (2015).
\bibitem{mu26} B. Gaveau, M. Moreau and  L. S. Schulman, Phys. Rev. E {\bf 82}, 051109 (2010).
\bibitem{mu27} E. Boukobza and D.J. Tannor,  Phys. Rev. Lett. {\bf 98}, 240601 (2007).
\bibitem{mu28} T. Bameta, D.  Das, R.  Padinhateeri and M. M. Inamdar,  ArXiv:1503.06529 (2015).
\bibitem{mu29} D. Oriola and J. Casademunt,  Phys. Rev. Lett. {\bf 111}, 048103 (2013).
\bibitem{mu30} O. Campa, Y. Kafri, K.B. Zeldovich, J. Casademunt and J.-F. Joanny, Phys. Rev. Lett. {\bf 97}, 038101 (2006).
\bibitem{muuu17} M. A. Taye,  Phys. Rev. E {\bf 94},  032111 (2016).
\bibitem{c1}
G.~A.~L.~Forão, F.~S.~Filho, B.~A.~N.~Akasaki, and C.~E.~Fiore,
Phys.\ Rev.\ E {\bf 110}, 054125 (2024).
\bibitem{c2}
I.~N.~Mamede, P.~E.~Harunari, B.~A.~N.~Akasaki, K.~Proesmans, and C.~E.~Fiore,
Phys.\ Rev.\ E {\bf 105}(2), 024106 (2022).
\bibitem{c3}
B.~A.~N.~Akasaki, M.~J.~de~Oliveira, and C.~E.~Fiore,
Phys.\ Rev.\ E {\bf 101}, 012132 (2020).
\bibitem{c4}
C.~E.~Fiore and M.~J.~de~Oliveira,
Phys.\ Rev.\ E {\bf 99}(5), 052131 (2019).
\bibitem{c5}
K.~Proesmans and C.~Van~den~Broeck,
Chaos {\bf 27}(10), 104601 (2017).
\bibitem{c6}
S.~Giordano, G.~Bonfanti, and C.~Van~den~Broeck,  % substitute actual authors
Phys.\ Rev.\ E {\bf 103}, 052116 (2021).
\bibitem{c7}
L.~Defaveri, C.~Olivares, and C.~Anteneodo,
Phys.\ Rev.\ E {\bf 105}(5), 054149 (2022).
\bibitem{muuu177} M. A. Taye,  Phys. Rev. E {\bf 101},  012131 (2020). 
\bibitem{mar2} H. Ge, Phys. Rev. E {\bf 89}, 022127 (2014).
\bibitem{mar1} H. K. Lee, C. Kwon, and H. Park, Phys. Rev. Lett. {\bf 110}, 050602 (2013).
\bibitem{mg1} T.~V. Vu, \textit{arXiv preprint arXiv:2411.19546} (2025).
\bibitem{mg2} K.-W. Kim, E.~Kwon, and Y.~Baek, \textit{arXiv preprint arXiv:2503.16958} (2025).
\bibitem{mg3} F.~J. Cao, \textit{Quantum BioSystems}, \textbf{7}, 75--90 (2025).
\bibitem{mg4} A.~del Campo, J.~Goold, and M.~Paternostro, \textit{Sci. Rep.}, \textbf{4}, 1--7 (2020).
\bibitem{mg5}
C.~E.~Fern\'andez\,Noa, P.~E.~Harunari, M.~J.~de~Oliveira, and C.~E.~Fiore,
Phys.\ Rev.\ E {\bf 100}, 012104 (2019).


\bibitem{mg6} S.~Ro et al., \textit{arXiv preprint arXiv:2105.12707} (2021).
\bibitem{am3} J. M. Sancho, M. S. Miguel and D. Duerr,  J. Stat. Phys. {\bf 28}, 291  (1982).
\bibitem{am33} A. M. Jayannavar and M. C. Mahato,  Pramana J. Phys. {\bf 45}, 369, (1995).
\bibitem{am4} K. Sekimoto,  J. Phys. Soc. Jpn.  {\bf 66}, 1234 (1997).
\bibitem{am5}K. Sekimoto,  Prog. Theor. Phys. Suppl. {\bf 130}, 17 (1998).
\bibitem{c8}
S.~Hottovy, A.~McDaniel, G.~Volpe, and J.~Wehr,
Commun.\ Math.\ Phys.\ {\bf 336}, 1259 (2015).
\bibitem{mar12} J. Bauermann  and  B. Lindner, Biosystems {\bf 178},25 (2019).
\bibitem{mar10} O.~Carrillo, M.~Ibáñez, J.~García-Ojalvo, J.~Casademunt, and J.~M.~Sancho, Phys.\ Rev.\ E {\bf 67}, 046110 (2003).
\bibitem{mar15} O.~Reynolds, Phil.\ Trans.\ R.\ Soc.\ Lond.\ {\bf 177}, 157 (1886).







%ibitem{mu23} J. Parrondo, B. Jimenez de Cisneros and R. Brito,  Stochastic Processes in Physics, Chemistry and Biology LNP557 (Springer-Verlag, Berlin (2000)), p. 38.
 %bibitem{mu40} S. Lahiri, S. Rana, and A. M. Jayannavar, Phys. Letters A, {\bf 378}, 979 (2014).

\end{thebibliography}
\end{document}